\documentstyle[epsfig,prl,aps]{revtex}

\begin{document}

\draft

\twocolumn

\title{On the universality of small scale turbulence}

\author{Ch.  Renner\textsuperscript{1}, J. 
Peinke\textsuperscript{1}\thanks{corresponding author; email: 
peinke@uni-oldenburg.de},
R. Friedrich\textsuperscript{2}, O. Chanal\textsuperscript{3} and B.
Chabaud\textsuperscript{3} }

\address{ \textsuperscript{1}Fachbereich Physik, Universit\"at
Oldenburg, D--26123 Oldenburg \\ \textsuperscript{2} Institut f\"ur
theoretische Physik, Universit\"at Stuttgart, D--70550 Stuttgart \\
\textsuperscript{3}CNRS-CRTBT, Universite Joseph Fourier, Grenoble,
France }

\date{\today} 
\maketitle
\begin{abstract}
    The proposed universality of small scale turbulence is
    investigated for a set of measurements in a cryogenic free jet
    with a variation of the Reynolds number (Re) from $8500$ to $10^6$
    (max $R_{\lambda} \approx 1200$).  The traditional analysis of the
    statistics of velocity increments by means of structure functions
    or probability density functions is replaced by a new method which
    is based on the theory of stochastic Markovian processes.  It
    gives access to a more complete characterization by means of joint
    probabilities of finding velocity increments at several scales. 
    Based on this more precise method our results call in question the
    concept of universality.
\end{abstract}


The complex behaviour of turbulent fluid motion has been the subject
of numerous investigations over the last 60 years and still the
problem is not solved \cite{turb}. Especially the unexpected frequent
occurences of high values for velocity fluctuations on small scales,
known as small scale intermittency, remain a challenging subject for
further investigations.

Following an idea by Richardson \cite{Richardson} and the theories by
Kolmogorov and Oboukhov \cite{K41,K62}, turbulence is usually assumed
to be universal in the sense that for scales $r$ within the inertial
range $\eta \ll r \ll L$ the statistics of the velocity field is
independent of the large scale boundary conditions, the mechanism of
energy dissipation and the Reynolds number (Re).  Here, $L$ denotes
the integral length scale and $\eta$ the dissipation length.

Besides its physical impacts, the assumed universality of the
turbulent cascade has gained considerable importance for models and
numerical methods such as large eddy simulations (LES), cf. 
\cite{LESZitat}.  Finding experimental evidence for the validity of
the assumed universality is therefore of utmost importance.

The turbulent cascade is usually investigated by means of the velocity
difference on a certain length scale $r$, the so-called longitudinal
velocity increment $u(r)$
\begin{equation}
	u(r) = {\mathbf e} \cdot \left[ {\mathbf v} \left( 
	{\mathbf x} + {\mathbf e} r,t \right) - {\mathbf v}
	\left( {\mathbf x},t \right) \right],
\end{equation}
where ${\mathbf v}$ and ${\mathbf e}$ denote the velocity and an unit
vector with arbitrary direction, respectively. Traditionally, the
statistics of $u(r)$ is characterized by its moments
$S_{u}^{n}(r)=\left< u^{n}(r) \right>$, the so-called structure
functions.  For scales $r$ within the inertial range, the structure
functions are commonly approximated by power laws in $r$:
$S_{u}^{n}(r) \propto r^{\zeta_{n}}$.  More pronounced scaling
behaviour is found for the so-called extended selfsimilarity method
\cite{ESS}.

Experimental investigations carried out in several flow configurations
at a large variety of Reynolds numbers yield strong evidence that the
scaling exponents $\zeta_{n}$ in fact show universal behaviour,
independent of the experimental setup \cite{ZetaNundRe}. A different
result, however, was found for the probability density functions (pdf)
$p(u,r)$. Recent studies using the
theoretical framework of infinitely divisible multiplicative cascades
show that the relevant parameters describing intermittency strongly
depend on the Reynolds number \cite{MalecotKahalerrasCastaing}.

From the point of view of statistics, a characterization of the scale
dependent disorder of turbulence by means of structure functions or
pdfs $p(u,r)$ is incomplete.  Theoretical studies \cite{Procaccia}
point out that a complete statistical characterization of the
turbulent cascade has to take into account the joint statistical
properties of several increments on different length scales.  An
experimental study concerned with the statistical properties of small
scale turbulence and its possible universalities therefore requires an
analyzing tool which is not based on any assumption on the underlying
physical process and which is capable of describing the multiscale
statistics of velocity increments.  Such a tool is given by the
mathematical framework of Markov processes.  Recently, it has been
shown that this tool allows to derive the stochastic differential
equations governing the evolution of the velocity increment $u$ in the
scale parameter $r$ from experimental data \cite{MarkovZitat,JFM}.

In this letter we present, firstly, our new method to analyse
experimental data, secondly, results for different Re-numbers,
thirdly, experimental findings which question the proposed
universality.

The stochastic process governing the scale dependence of the velocity
increment is Markovian, if the conditional pdf
$p(u_{1},r_{1}|u_{2},r_{2};...; u_{N},r_{N})$ fulfills the relation
\cite{Risken,K31}:
\begin{equation}
	p(u_{1},r_{1}|u_{2},r_{2};...;u_{N},r_{N}) 
	= p(u_{1},r_{1}|u_{2},r_{2}) . \label{MarkovDef}
\end{equation}
The conditional pdf $p(u_{1},r_{1}|u_{2},r_{2};...; u_{N},r_{N})$
describes the probability for finding the increment $u_{1}$ on the
smallest scale $r_{1}$ provided that the increments $u_{2},...,u_{N}$
are given at the larger scales $r_{2},...,r_{N}$.  We use the
conventions $r_{i} \leq r_{i+1}$ and $u_{i}=u(r_{i})$.  It could be
shown in \cite{JFM,EPL} that experimental data satisfy equation
(\ref{MarkovDef}) for scales $r_{i}$ and differences of scales $\Delta
r = r_{i+1}-r_{i}$ larger than an elementary step size $l_{mar}$, 
comparable to the mean free path of molecules undergoing a Brownian
motion.

As a consequence of (\ref{MarkovDef}), the joint pdf of $N$ increments
on $N$ different scales simplifies to:
\begin{eqnarray}
	p(u_{1},r_{1};u_{2},r_{2};...;u_{N},r_{N}) =
	p(u_{1},r_{1}|u_{2},r_{2}) \times \nonumber \\
	\times p(u_{2},r_{2}|u_{3},r_{3})...
	p(u_{N-1},r_{N-1}|u_{N},r_{N}) p(u_{N},r_{N}) .
	\label{chain}
\end{eqnarray}
Equation (\ref{chain}) indicates the importance of the Markovian
property for the analysis of the turbulent cascade: The entire
information about the stochastic process, i.e. any $N$--point or, to
be more precise, any $N$--scale distribution of the velocity
increment, is encoded in the conditional pdf $p(u,r|u_{0},r_{0})$
(with $r\leq r_{0}$). 

Furthermore, it is well known that for Markovian processes the
evolution of $p(u,r|u_{0},r_{0})$ in $r$ is described by the
Kramers--Moyal--expansion \cite{Risken}.  For turbulent data it was
verified that this expansion stops after the second term \cite{JFM}. 
Thus the conditional pdf $p(u,r|u_{0},r_{0})$ is described by the
Fokker-Planck equation:
\begin{eqnarray}
	-r\frac{\partial}{\partial r} p(u,r|u_{0},r_{0})  =  
	- \frac{\partial}{\partial u} && \left( D^{(1)}(u,r) 
	p(u,r|u_{0},r_{0}) \right) \nonumber \\
	 + \frac{\partial}{\partial u^2} && \left(  
	D^{(2)}(u,r) p(u,r|u_{0},r_{0}) \right) . 
	\label{FoPlaCond}
\end{eqnarray}
By multiplication with $p(u_{0},r_{0})$ and integration with respect
to $u_{0}$, it can be shown that the single scale pdf $p(u,r)$ obeys
the same equation.

Another important feature of the Markov analysis is the fact that the
coefficients $D^{(1)}$ and $D^{(2)}$ (drift and diffusion coefficient,
respectively) can be extracted from experimental data in a parameter 
free way by their mathematical definition, see \cite{Risken,K31}:
\begin{eqnarray}
	D^{(k)}(u,r) & = & \lim_{\Delta r \rightarrow 0}
	\frac{r}{k! \Delta r} M^{(k)}(u,r,\Delta r), 
	\label{DnDef} \\
	M^{(k)}(u,r,\Delta r) & = & 
	\int\limits_{-\infty}^{+\infty} ( \tilde{u}-u)^{k} 
	p(\tilde{u},r-\Delta r | u,r) d\tilde{u}.
	\label{MnDef}
\end{eqnarray}
The conditional moments $M^{(k)}(u,r,\Delta r)$ can easily be
calculated from experimental data. Approximating the limit $\Delta r
\rightarrow 0$ in equation (\ref{DnDef}) by linear extrapolation then
yields estimates for the $D^{(k)}(u,r)$.

As a next point, we focus on the analysis of experimental data
measured in a cryogenic axisymmetric helium gas jet at Reynolds
numbers ranging from $8500$ to $757000$.  Each data set contains $1.6
\cdot 10^7$ samples of the local velocity measured in the center of
the jet in a vertical distance of $40 D$ from the nozzle using a
selfmade hotwire anemometer ($D=2mm$ is the diameter of the nozzle). 
Taylor's hypothesis of frozen turbulence was used to convert time lags
into spatial displacements.  Following the convention chosen in
\cite{JFM}, the velocity increments for each data set are given in
units of $\sigma_{L}= \sqrt{2} \sigma$, where $\sigma$ is the standard
deviation of the velocity fluctuations of the respective data set.

\begin{figure}[]
  \begin{center}
    \epsfig{file=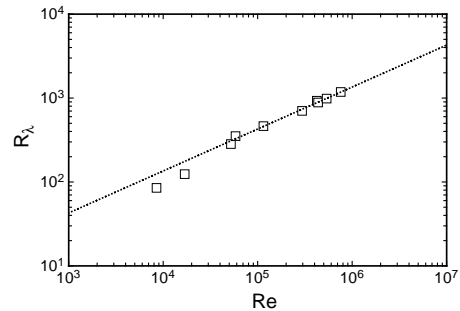, width=6.0cm}
  \end{center}
 \caption{ \it Taylor--scale Reynolds number $R_{\lambda}$ (for
 details on the determination see [11]) as a function of the
 nozzle-based Reynolds number $Re$.  Dotted line: $R_{\lambda} = 1.35
 \sqrt{Re}$.  }
 \label{ReLamVonRe}
\end{figure}

In order to check consistency of the data with commonly accepted
features of fully developed turbulence, we calculated the dependence
of the Taylor--scale Reynolds number $R_{\lambda}$ on the nozzle-based
Reynolds number. Figure \ref{ReLamVonRe} shows that $R_{\lambda}$
scales like the square root of $Re$, in accordance with theoretical
considerations and earlier experimental results.  Further
details on the experimental setup can be found in \cite{Chanal}.
%
%
The condition (\ref{MarkovDef}) for the Markov property was checked
using the method proposed in ref.  \cite{JFM}.  For all the data sets,
the Markovian property was found to be valid for scales $r_{i}$ and
differences of scales $\Delta r = r_{i+1}-r_{i}$ larger than the
elementary step size $l_{mar}$, which turned out to be of the order of
magnitude of the Taylor microscale $\lambda$ for all $Re$--numbers
investigated.

Having determined the Markov length $l_{mar}$, the coefficients
$D^{(1)}(u,r)$ and $D^{(2)}(u,r)$ can be estimated from the measured
conditional moments $M^{(1)}$ and $M^{(2)}$ according to 
equation (\ref{DnDef}). The extrapolation towards $\Delta r = 0$ was
performed fitting linear functions to the measured $M^{(k)}$ in the
intervall $l_{mar} \leq \Delta r \leq 2 l_{mar}$
\cite{ExtrapolComment}.

\begin{figure}[]
    \begin{center}
       \epsfig{file=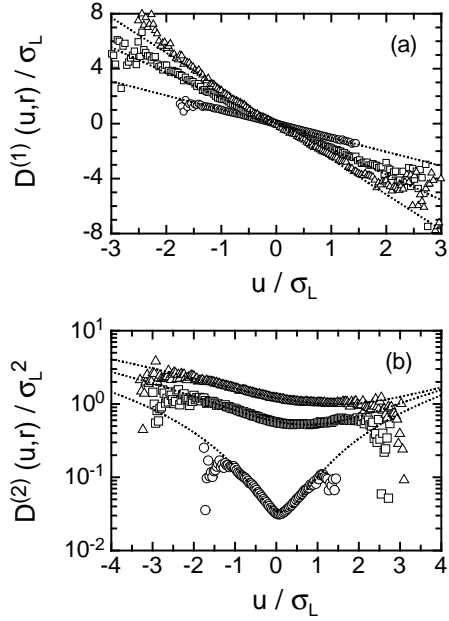,  width=6.0cm }
     \end{center}
 \caption{ \it Coefficients $D^{(1)}(u,r)$ (a) and $D^{(2)}(u,r)$ (b)
 as functions of the velocity increment $u$ at $r=3\lambda$ (circles),
 $r=L/2$ (squares) and $r=L$ (triangles).  The dotted curves
 correspond to linear (a) and polyonmial (b) (degree two) fits to the
 measured data.  }
 \label{D12vonU}
\end{figure}

Figure \ref{D12vonU} shows the resulting estimates for the
coefficients $D^{(1)}$ and $D^{(2)}$ for the data set at
$R_{\lambda}=1180$ as a function of the velocity increment at several
scales $r$. The coefficients exhibit linear and quadratic
dependencies on the velocity increment, respectively:
\begin{eqnarray}
	D^{(1)}(u,r) & = & - \gamma(r) u , \nonumber \\
	D^{(2)}(u,r) & = & \alpha(r) - \delta(r) u + \beta(r) u^2.
	\label{DkFunktional}
\end{eqnarray}
Equation (\ref{DkFunktional}) is found to describe the dependence of
the $D^{(k)}$ on $u$ for all scales $r$ as well as for all Reynolds
numbers investigated. By fitting the coefficients $D^{(k)}$ according
to (\ref{DkFunktional}), it is thus possible to determine the scale
dependence of the coefficients $\gamma$, $\alpha$, $\delta$ and
$\beta$. 

\begin{figure}[]
  \begin{center}
    \epsfig{file=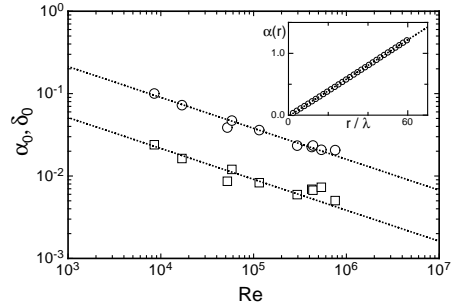, width=6.0cm}
  \end{center}
 \caption{ \it Coefficients $\alpha_{0}$ (circles) and $\delta_{0}$
 (squares) defined in eqs.  (\ref{DkFunktional}) and (\ref{AundBvonR})
 as functions of the Reynolds number $Re$; dotted lines represent
 power laws in $Re$ with an scaling exponent of $-3/8$.  The inlet
 displays $\alpha(r)$ as a function of the length scale $r$ for
 $R_{\lambda} = 1180$.  }
 \label{AlphaDeltaNull}
\end{figure}

The constant and linear coefficient of $D^{(2)}$, $\alpha$ and
$\delta$, turn out to be linear functions of the scale $r$ (see the 
inlet in fig. \ref{AlphaDeltaNull}):
\begin{eqnarray}
	\alpha(r) = \alpha_{0} \frac{r}{\lambda} , \qquad
	\delta(r) = \delta_{0} \frac{r}{\lambda} . \label{AundBvonR}
\end{eqnarray}
%
%

As shown in figure \ref{AlphaDeltaNull}, the slopes $\alpha_{0}$ and
$\delta_{0}$ show strong dependencies on the Reynolds number and can
be described by power laws in $Re$ with an scaling exponent of $-3/8$:
\begin{eqnarray}
	\alpha_{0} \approx 2.8 Re^{-3/8} , \qquad 
	\delta_{0} \approx 0.68 Re^{-3/8} . \label{ADNullVonRe}
\end{eqnarray}

%
%

\begin{figure}[]
  \begin{center}
    \epsfig{file=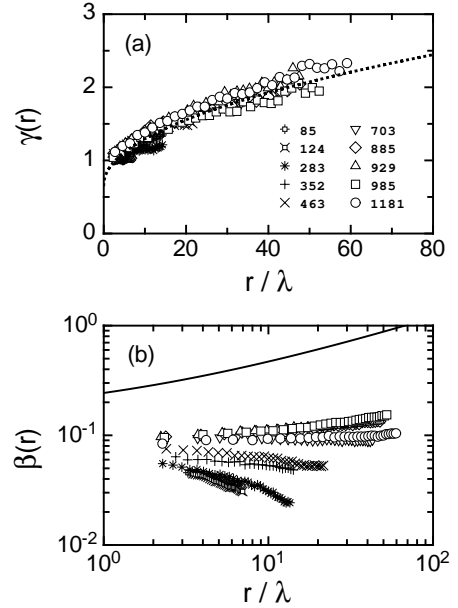, width=6.0cm}
  \end{center}
 \caption{ \it The slope $\gamma(r)$ of $D^{(1)}$ (a) and the
 quadratic coefficient $\beta(r)$ of $D^{(2)}$ (b) as functions of the
 scale $r$ for several Reynolds numbers (see legend).  $\gamma$ is
 close to a universal function of the scale $\rho=r/\lambda$ (the
 dotted line is a fit according to eq.  (\ref{GammaVonR})), the
 coefficient $\beta$ exhibits a strong dependence on the Reynolds
 number with a clear tendency towards the limiting value
 $\beta_{\infty}(r)$ given by equ.  (\ref{BetaInfty}) (full line). 
 }
 \label{GammaUndBeta}
\end{figure}

A different result is obtained for the linear term $\gamma(r)$ of
$D^{(1)}$, see figure \ref{GammaUndBeta}. It turns out to be a
universal function of $r/\lambda$ and is found to be well described by
\begin{equation}
	\gamma(r) = \frac{2}{3} + c \sqrt{\frac{r}{\lambda}}, \label{GammaVonR}
\end{equation}
where $c=0.20 \pm 0.01$. 

%
%

These results allow for a statement on the limiting case
of infinite Reynolds numbers, $Re \rightarrow \infty$. According to eq. 
(\ref{ADNullVonRe}) the coefficients $\alpha$ and $\delta$ tend to
zero \cite{CommentII}. $\gamma(r)$ does not
depend on $Re$. Thus drift- and diffusion coefficient take the following
simple form for $Re \rightarrow \infty$:
\begin{eqnarray}
	D^{(1)}_{\infty}(u,r) & = & - \gamma(r) u , \nonumber \\
	D^{(2)}_{\infty}(u,r) & = & \beta_{\infty}(r) u^2 . 
	\label{D12Infty}
\end{eqnarray}
Based on this limiting result we discuss next implications for the
structure functions $S_{u}^{n}(r)$. After the multiplication of the
corresponding Fokker--Planck equation (\ref{FoPlaCond}) for $p(u,r)$
with $u^n$ from left and successively integrating with respect to $u$,
the equation
\begin{equation}
	\frac{r \frac{\partial}{\partial r}S_{u}^{n}(r) }
	{ n S_{u}^{n}(r) } = \gamma(r) - (n-1) \beta_{\infty}(r)
	\label{SunInftyGlg}
\end{equation}
is obtained. 

According to Kolmogorov's four-fifth law, cf.  \cite{turb}, the third
order structure function, $S_{u}^{3}(r)$, is proportional to $r$. 
Thus, for $n=3$, the  left side of eq. (\ref{SunInftyGlg}) is equal
to $1/3$ and $\beta_{\infty}(r)$ is given by:
\begin{eqnarray}
	\beta_{\infty}(r) = \frac{\gamma(r)}{2} - \frac{1}{6}.
	\label{BetaInfty}
\end{eqnarray}
For increasing Reynolds numbers, the experimental results for
$\beta(r)$ in fact show a tendency towards the limiting value
$\beta_{\infty}$ as given by eq.  (\ref{BetaInfty}) (see fig. 
\ref{GammaUndBeta}), but it is also clearly observed that the convergence
is slow and that even the highest accessible Reynolds numbers are still
far from this limiting case.

To summarize, the mathematical framework of Markov processes can
succesfully be applied to characterize the stochastic behaviour of
turbulence with increasing Reynolds number.  Moreover, the description
obtained by our method is complete in the sense that the entire
information about the stochastic process, i.e. the information about
any $N$--scale pdf $p(u_{1},r_{1};u_{2}, r_{2};...;u_{N},r_{N})$, is
encoded in the two coefficients $D^{(1)}(u,r)$ and $D^{(2)}(u,r)$, for
which we find rather simple dependencies on their arguments $u$, $r$
and the Reynolds number.

The $Re$--dependence of the coefficients, especially of $D^{(2)}$,
yields strong experimental evidence for a significant change of the
stochastic process as the Reynolds number increases.  This finding
clearly contradicts the concept of a universal turbulent cascade and
might also be of importance in large eddy simulations where the
influence of the subgrid stress on the large scale dynamics of a
turbulent flow is modeled under the assumption of universality.

It is easily verified that, according to eq.  (\ref{SunInftyGlg}), the
increase of $\beta(r)$ with $Re$ excludes the simple scaling laws
proposed by Kolmogorov in 1941 \cite{K41} even for $Re \rightarrow
\infty$.  Furthermore, the universal functional dependence of
$\gamma(r)$ on $r$ (eq.  (\ref{GammaVonR})) does not support the
recently proposed constant value of $\gamma \approx 1/3$
\cite{MarkovZitat,YakhotDavoudi}.  The obvious dependence of the
coefficients $\gamma$ and $\beta$ on $r$ also contradicts the
assumption that the structure functions exhibit scaling behaviour for
all orders $n$, as can be derived from eq.  (\ref{SunInftyGlg}).

With the limiting values for the coefficients $D^{(k)}$ as given by
eq.  (\ref{D12Infty}), the stochastic process for infinite Reynolds
numbers corresponds to an infinitely divisible multiplicativ cascade
\cite{AmblardBrossier} as proposed in ref.  \cite{CastaingZitat}. 
However, from the slow convergence of the measured coefficient 
$\beta(r)$ towards its limiting value $\beta_{\infty}(r)$, it is
obvious that turbulent data measured in typical laboratory experiments
are still far from that special case.  It therefore seems questionable
to us whether models on turbulence established under the assumption of
infinite Reynolds numbers can be tested in real-life experimental
situations at all.

Acknowledgement: We gratefully acknowledge fruitful discussions with 
J.-F. Pinton, B. Castaing, F. Chilla and M. Siefert.


\begin{references}{}
\bibitem{ExtrapolComment} Note that this method differs slightly from
the one proposed in ref.  \cite{JFM}, where first the $M^{(k)}$ were
fitted as functions of the velocity increment $u$.  The coefficients
of those fits were extrapolated towards $\Delta r = 0$ in a second
step.
\bibitem{CommentII} Strictly speaking, only the coefficients
$\alpha_{0}$ nd $\delta_{0}$ tend to zero.  But as the ratio
$L/\lambda$ scales like $R_{\lambda}$, the coefficients $\alpha$ and
$\delta$ at the integral length scale $L$ like: $\alpha(L) =
\alpha_{0} \frac{L}{\lambda} \propto Re^{-3/8} R_{\lambda} \propto
R_{\lambda}^{-3/4} R_{\lambda} \propto R_{\lambda}^{1/4}$.  On the
other hand, the coefficient $\gamma$ scales like the square root of
$r/\lambda$.  Thus: $\gamma(L) \propto \sqrt{L/\lambda} \propto
R_{\lambda}^{1/2}$.  The fact that $\gamma(L)$ grows faster than
$\alpha(L)$ and $\delta(L)$ justifies neglecting $\alpha$ and $\delta$
in the limit $ Re \rightarrow \infty $.
\end{references}
\end{document}